\documentclass[apj]{emulateapj}
\usepackage{apjfonts}
\usepackage{float}
\usepackage{booktabs}
\usepackage{amsmath}
\shorttitle{Dust-Obscured Bulge Building}
\shortauthors{Nelson et al.}
\slugcomment{Submitted to ApJ}

\begin{document}

\gdef\ha{H$\alpha$}
\gdef\nii{[N\,{\sc\,ii}]}
\gdef\h{$H_{F160W}$}
\gdef\j{$J_{F125W}$}
\gdef\gn{GOODSN-18574}
\gdef\restlam{rest-500$\mu$m}
\gdef\msun{M$_{\odot}$}
\gdef\m{M$_*$}
\gdef\sigmaone{$\Sigma_1$}
\gdef\sigmacen{$\Sigma_{cen}$}
\gdef\deltam{$\Delta$M$_*$}
\gdef\mic{$\mu$m}
\gdef\aha{$A_{H\alpha}$}
\gdef\betas{$\beta-$slope}
\gdef\ebv{$E(B-V)$}
\gdef\auv{$A_{UV}$}
\gdef\uv{\textit{uv}}
\gdef\prog{Andromeda progenitor}

\title{Millimeter mapping at $\lowercase{z}\sim1$: dust-obscured bulge building and disk growth}
\author{Erica J. Nelson\altaffilmark{1,2},
Ken-ichi Tadaki\altaffilmark{3},
Linda J. Tacconi\altaffilmark{1},
Dieter Lutz\altaffilmark{1},
Natascha M. F\"orster Schreiber\altaffilmark{1},
Anna Cibinel\altaffilmark{4},
Stijn Wuyts\altaffilmark{5},
Philipp Lang\altaffilmark{6},
Mireia Montes\altaffilmark{7,8},
Pascal A. Oesch\altaffilmark{9},
Sirio Belli\altaffilmark{1},
Rebecca L. Davies\altaffilmark{1},
Richard I. Davies\altaffilmark{1},
Reinhard Genzel\altaffilmark{1},
Magdalena Lippa\altaffilmark{1},
Sedona H. Price\altaffilmark{1},
Hannah \"Ubler\altaffilmark{1},
Emily Wisnioski\altaffilmark{10}
}

\altaffiltext{1}{Max-Planck-Institut f\"ur extraterrestrische Physik,
Giessenbachstrasse, D-85748 Garching, Germany}
\altaffiltext{2}{Harvard-Smithsonian Center for Astrophysics, 60 Garden St, 
Cambridge, MA 02138, USA}
\altaffiltext{3}{National Astronomical Observatory of Japan, 2-21-1 Osawa, Mitaka, Tokyo 181-8588, Japan}
\altaffiltext{4}{Astronomy Centre, Department of Physics and Astronomy, University of Sussex, Brighton, BN1 9QH, UK}
\altaffiltext{5}{Department of Physics, University of Bath, Claverton Down, Bath, BA2 7AY, UK}
\altaffiltext{6}{Max Planck Institute for Astronomy (MPIA), K\"nigstuhl 17, 69117, Heidelberg, Germany}
\altaffiltext{7}{School of Physics, University of New South Wales, Sydney, NSW 2052, Australia}
\altaffiltext{8}{Astronomy Department, Yale University, New Haven, CT 06511, USA}
\altaffiltext{9}{Observatoire de Gen\`eve, 51 Ch. des Maillettes, 1290 Versoix, Switzerland}
\altaffiltext{10}{Research School of Astronomy \& Astrophysics, Australian National University, Canberra, ACT-2611, Australia}

\begin{abstract}
A randomly chosen star in today's Universe is most likely to live in a galaxy with a 
stellar~mass between that of the Milky Way and Andromeda. 
Yet it remains uncertain how the structural evolution of these bulge-disk
systems proceeded. Most of the unobscured star formation we 
observe building \prog s at 
0.7\,$<$\,$z$\,$<$\,1.5 occurs in disks, but $\gtrsim90\%$ of their 
star formation is reprocessed by dust and remains unaccounted for. 
Here we map \restlam\ dust continuum emission in an \prog\ 
at $z=1.25$ to probe where it is growing through dust-obscured star formation. 
Combining resolved dust measurements from the NOEMA interferometer 
with Hubble Space Telescope \ha\ maps and multicolor imaging 
(including new UV data from the HDUV survey), we find 
a bulge growing by dust-obscured star formation:
while the unobscured star formation is centrally suppressed, 
the dust continuum is centrally concentrated, filling in the ring-like structures
evident in the \ha\ and UV emission.
Reflecting this, the dust emission is more compact than the 
optical/UV tracers of star formation with
$r_e(dust)=3.4$kpc, $r_e(H\alpha)/r_e(dust)=1.4$, and $r_e(UV)/r_e(dust)=1.8$.
Crucially, however, the bulge and disk of this galaxy are building simultaneously;
although the dust emission is more compact than the 
rest-optical emission ($r_e(optical)/r_e(dust)=1.4$), it is somewhat
less compact than the stellar mass ($r_e(M_*)/r_e(dust)=0.9$).
Taking the \restlam\ emission as a tracer of star formation, 
the expected structural evolution of this galaxy can be accounted 
for by star formation: it will grow in size by $\Delta r_e/\Delta M_*\sim0.3$
and central surface density by $\Delta \Sigma_{cen}/\Delta M_*\sim0.9$.
Finally, our observations are consistent with a 
picture in which merging and disk 
instabilities drive gas to the center of galaxies, boosting global star 
formation rates above the main sequence and building bulges. 

\end{abstract}


\section{Introduction}

Owing to significant investments in optical and near-infrared instrumentation, 
we now have high resolution mapping of large numbers galaxies in the 
rest-UV+optical during the epoch when they formed most of their stars ($1<z<3$).
This mapping has shown that most star formation as traced by \ha\ and UV emission
occurs in clumpy, rotating galactic disks 
\citep[e.g.][]{Genzel:06,Genzel:08,Forster-Schreiber:06,Forster-Schreiber:09,
Wisnioski:11,Kassin:12,Wuyts:13,Nelson:13,Wisnioski:15,Stott:16}.
Additionally, studies mapping where galaxies are growing via rest-UV/optical
tracers of the specific star formation rate ($sSFR=SFR/M_*$) which trace
present star formation relative to the integral of past star formation 
(e.g. EW(\ha), UV-optical color gradients) 
typically find either flat or somewhat centrally depressed 
sSFR, meaning galaxies are generally growing somewhere between self-similarly,
and inside-out; not, on average, getting more compact
\citep{Wuyts:12,Nelson:12,Nelson:13,Nelson:16b,Liu:16,Liu:17}. 
However, a significant fraction of star formation is attenuated by dust and may be 
missed by these types of observations. 
Most importantly, this hampers our ability 
to determine where within galaxies most of the stars were formed
and consequently how galaxies grew through star formation.

\begin{figure*}
\centering
\includegraphics[width=0.7\textwidth]{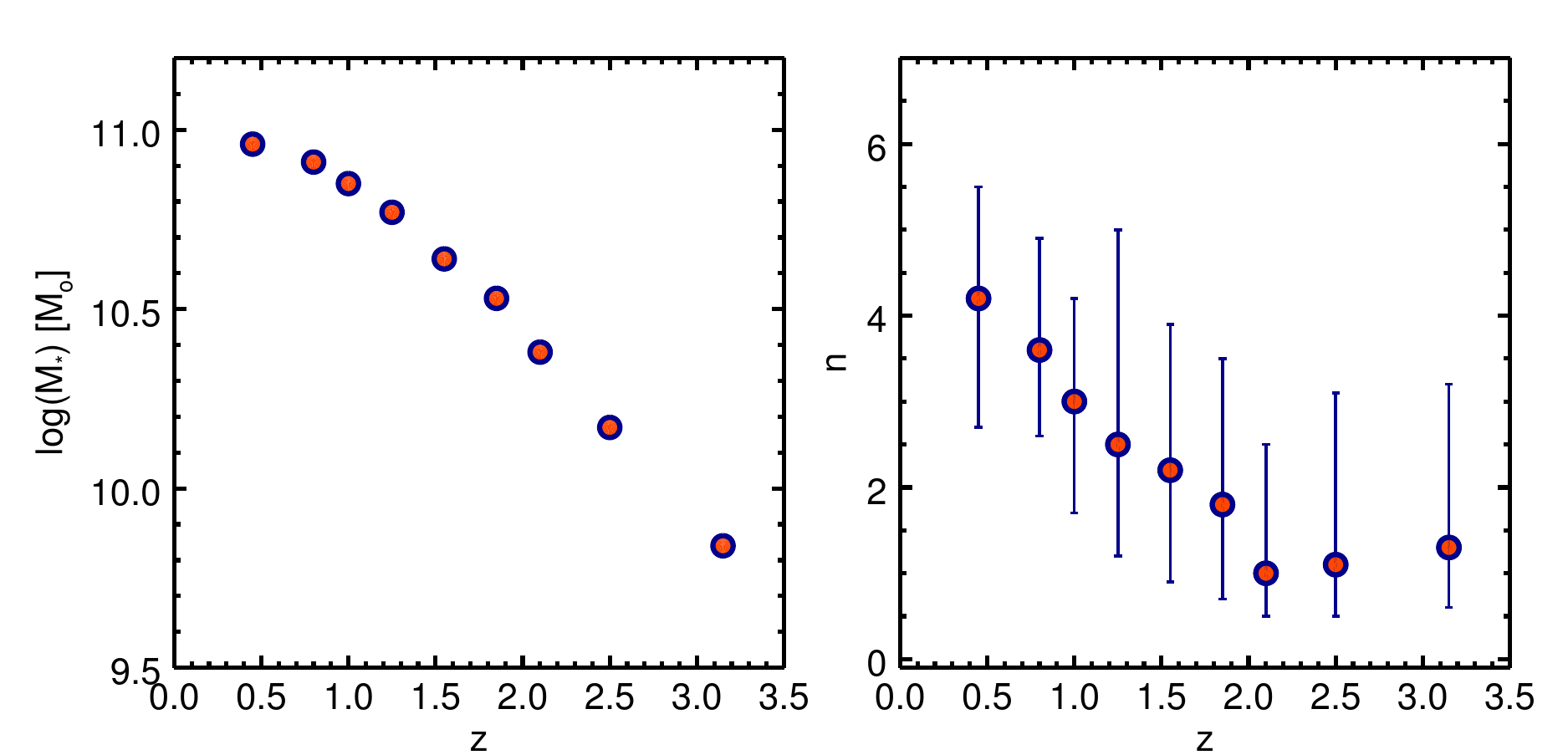}
\caption{ Left: Average stellar mass evolution of galaxies with present-day 
    mass of the Andromeda based on abundance matching 
    \citep{Moster:13, Behroozi:13}.  
    In this paper we select a galaxy that based on its stellar mass is likely 
    to be a $z\sim1$ progenitor of a galaxy with present-day mass similar to Andromeda
    \citep[e.g.][]{Papovich:15}. 
    Right: The evolution of the \h\ S{\'e}rsic indices with time \citep{Papovich:15}.
    The steady increase in the S{\'e}rsic indices during the epoch of this study suggests that 
    this is a critical epoch for understanding the growth of galactic bulges.  
\label{fig:selection}
}
\end{figure*}

Recent studies of the spatially-resolved Balmer decrements, colors, 
and spectral energy distributions (SEDs) of 
large samples of galaxies have found that with increasing stellar mass, 
both the normalization and gradient in dust attenuation increases 
\citep[e.g. ][]{Wuyts:12,Nelson:16a,Liu:16,Liu:17,Wang:17}.
At a most basic level, this suggests that the dust-obscured star formation
may be distributed differently than the unobscured star formation in 
massive galaxies. In particular, it may be more compact. 
In massive galaxies near the peak of the cosmic star formation history 
($M_*\gtrsim2\times10^{10}$\msun\ at $z\sim1-2$),
typically $\gtrsim90\%$ of the emission from star formation is absorbed by dust
and re-radiated in the infrared (IR)
\citep[e.g.][]{Reddy:06,Reddy:10,Whitaker:12,Wuyts:11a,Whitaker:14,Reddy:17}. 
Thus, to determine how galaxies are building,
 it is essential to be able to map not only the unobscured 
star formation but also the obscured star formation. 
This has been difficult because telescopes operating at the far-IR (FIR) wavelengths 
necessary to probe the peak of the dust emission have had 
insufficient sensitivity and spatial resolution to map galaxies at $z\sim1-3$. 

With the increased sensitivity and spatial resolution 
of mm/submm interferometers such as NOEMA and ALMA, 
we can now map the dust continuum emission at longer wavelengths,
however. 
For galaxies near the peak of the cosmic SFH at $1<z<3$, these
interferometers can be used to efficiently probe dust continuum emission 
at rest-wavelengths $\sim 200-500$\mic.
As the emission at these wavelengths represents thermal emission from dust 
largely heated by star formation for the rapidly star forming galaxies at this epoch,
 it has been used as tracer of obscured star 
formation (modulo dust temperature gradients) \citep[e.g.][]{Barro:16b,Tadaki:17a}. 
A number of individual galaxies have now been mapped at millimeter wavelengths
revealing in a significant fraction very centrally concentrated molecular gas and dust 
\citep[e.g.][]{Tacconi:08,Tacconi:10,Barro:16b,Tadaki:17a}. Very compact sizes have also
been found in bright sources at centimeter wavelengths (10GHz) \citep{Murphy:17}.
In these massive galaxies, while star formation as traced by \ha\ emission is 
in extended, rotating disks, the star formation inferred from dust emission is 
much more centrally concentrated, building their centers 
\citep{Genzel:13,Tadaki:17a}.
This suggests dust-obscured in-situ star formation could be an important formation 
channel for the dense cores of massive galaxies.  
With $M_*$\,$\sim$\,$10^{11}$\,$M_\odot$ at $z\sim2$,
these galaxies are likely to be the progenitors of today's massive elliptical galaxies.
The next key question is how dust-obscured star formation is distributed 
in the progenitors of today's M* galaxies at the equivalent epoch, pushing 
dust mapping from the most massive galaxies down to more typical galaxies.

It remains uncertain which processes are responsible for building bulges in local
massive spirals \citep[e.g.][]{Kormendy:16}.
Even for the closest, best studied examples of the Milky Way and Andromeda,
the fossil record (stellar ages, abundances, dynamics, and structural parameters)
points to a first rapid and dissipative formation event
followed by secular growth, but the mechanisms involved remain unclear
\citep[e.g.][]{Saglia:10,Courteau:11,Dorman:12,Bland-Hawthorn:16}. 
In recent years, theoretical considerations, numerical simulations,
and empirical results on the structure and kinematics of high-$z$
star-forming galaxies have brought forward new bulge formation channels
through efficient disk-internal dissipative processes in the typically
gas-rich and turbulent $z\sim2$ disks.
These $z\sim2$ disks typically have baryonic gas mass fractions 
of $\sim$\,30\%-50\% \citep[e.g.][]{Tacconi:13} and intrinsic gas velocity
dispersions $\sim25-50$\,km/s 
\citep{Genzel:06,Forster-Schreiber:06,Forster-Schreiber:09,Kassin:07,
Law:09,Epinat:09,SNewman:13,Wisnioski:15,Stott:16}.
In these gas rich-turbulent disks, processes like violent disk instabilities and 
inward migration of giant star-forming clumps
may even lead to ``classical'' bulges without the need for merger events
\citep[e.g.][]{Immeli:04,Genzel:08,Zolotov:15,Bournaud:16}. 
Though the importance of these processes is debated 
\citep[e.g.][]{vandokkum:15,Lilly:16}.
With evolving gas inflow rates, sizes, merger rates and surface densities of 
gas and stars, the physics of disks at $z\sim1$ may be very different.
In particular, is this how the bulges of M* galaxies in the local universe are built?
Do they have an equivalent central dust obscured star formation phase before 
quenching? 
The potentially complex bulge formation histories underscore the importance
of in-situ studies, at epochs when galaxies were most actively forming their stars.

\begin{figure*}
\centering
\includegraphics[width=\textwidth]{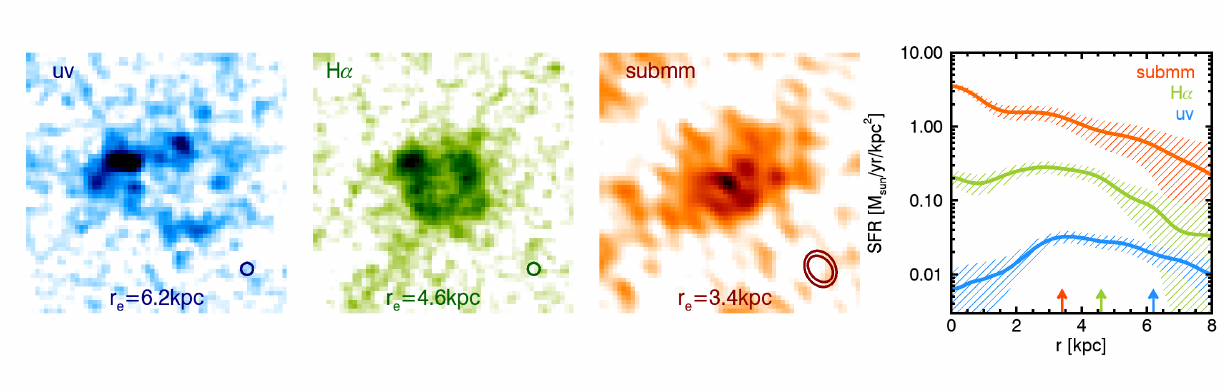}
\caption{Images and surface brightness profiles of the star formation in 
\gn\ as traced by rest-UV, \ha from HST and rest-submillimeter (\restlam) from 
NOEMA. 
The effective radius in each band is listed at the bottom of the image
and shown as an arrow in the plot of the radial profiles. 
It is clear from the images, surface brightness profiles, and radii that 
the different tracers of star formation trace very different regions.
The star formation gets more compact moving from less to more obscured
tracers. 
The UV is only seen at large radii, the \ha\ is somewhat more 
compact but still centrally depressed, while the submm is 
centrally concentrated. 
The \ha\ and even more so the UV, exhibit ring-like structures which 
are filled in by the dust-obscured star formation as traced by the submm.
The dark circles/ellipses in the bottom right corner show the FWHM resolution of the images. 
The shaded regions reflect the uncertainties due to the noise in the images.
\label{fig:imsandprofs}
}
\end{figure*}

Based on abundance matching arguments, we can link progenitor-descendant
populations across cosmic time 
\citep[e.g.][]{Conroy:09,vanDokkum:13,Behroozi:13,
 Moster:13,Papovich:15,Torrey:15,Torrey:17b,Wellons:17}. 
In this paper we use abundance matching to select a galaxy that based on its 
stellar mass is likely to have the same mass as Andromeda at $z=0$.
Throughout this paper, when we refer to this galaxy as an \prog, we mean 
that it is likely to be the progenitor of a galaxy which is part of the population 
of galaxies at $z=0$ which have the same mass as Andromeda.
The evolution of this population of galaxies inferred from abundance 
matching is shown in Fig.\ref{fig:selection}.  
Andromeda has a stellar mass of $M_*=1-1.5\times10^{11}$\msun, $\sim30\%$ 
of which is in the bulge \citep[e.g.][]{Geehan:06,Tamm:12}.
$z\sim1$ is a crucial epoch for studying bulge growth in these galaxies, with steadily 
increasing S{\'e}rsic indices suggesting significant structural evolution and 
bulge build-up \citep[e.g][]{vanDokkum:13,Lang:14,Papovich:15}. 
Probably related, the quenched fraction is also increasing during this epoch 
with the quiescent fraction amongst \prog\  increases from 
47\% at $z=1.4$ to 70\% at $z=0.7$ \citep{Papovich:15}.

In this paper, we combine new spatially-resolved 1.1mm (\restlam) 
data with Hubble Space Telescope \ha\ maps and UV-NIR imaging 
to investigate growth patterns in the progenitor of an \prog\ galaxy at $z=1.25$. 
This paper is organized as follows.
In \S2 we describe the target selection, the reduction and analysis of the 
NOEMA and HST data, and the derivation of spatially-resolved stellar population 
properties. 
In \S3 we discuss the derivation structural parameters of galaxy growth in \restlam, \ha,
UV, rest-optical continuum, and stellar mass. 
We compare the size, concentration, and radial profiles 
(SFR and sSFR) in the different
tracers as well as the effectiveness of an SED-based dust correction 
to the \ha\ and UV data. 
In \S4 we consider structural growth due to star formation in the context of 
the evolution of central density via the \sigmaone-\m\ relation and the expected
size evolution of Andromeda progenitors. 
Additionally, we compare the dust continuum size of our \prog\ to sizes measured 
for the progenitors of massive elliptical galaxies from \cite{Tadaki:17a}.

\section{Data}
\subsection{Selection}

The aim of this initiative was to map the sub-millimeter dust continuum emission 
in the progenitor of an M* galaxy during the time when it was likely to be building 
its bulge around $z\sim1$. 
Additionally, the availability of \ha\ maps at HST resolution for galaxies with 
0.7\,$<$\,$z$\,$<$\,1.5 allows for a direct comparison between 
the distribution of obscured and unobscured tracers
of star formation.
We selected galaxies with stellar masses between the expected stellar masses 
of Milky Way and Andromeda progenitors in this redshift range based on 
abundance matching \citep{Moster:13} (see Fig. \ref{fig:selection}). 
To facilitate our exploratory study with NOEMA, we wanted to target a galaxy for 
which we have sufficiently high S/N to accurately measure
the radial distribution and effective radius of the \ha\ emission for comparison to the 
\restlam\ data.
Finally, we required galaxies to have high $SFR(IR)>50$\,\msun/yr based on
Spitzer/MIPS and Herschel/PACS plus
$r_e>0.5$" to optimize detection and spatially resolved mapping with NOEMA.
Two galaxies were observed with NOEMA for relatively short integrations at 
low resolution then the one with the stronger detection was chosen for mapping.
Thus, the pilot target selected for this exploratory study was \gn.
This galaxy has
z=1.248, $M_*=6.76\times10^{10}M_\odot$, 
$SFR(IR,H\alpha,UV)=(164,28,5)$\msun/yr
\footnote{These \ha\ and UV SFRs are not dust-corrected.}, and
$r_e(H_{F160W})=0.56$",  
putting this galaxy roughly 
on the size-mass relation \citep{vanderWel:14} 
and $\sim0.55$dex above the SFR-mass relation at this redshift
\citep{Whitaker:14}. 

\subsection{NOEMA data reduction and analysis}\label{section:noemadata}

\begin{figure*}
\centering
\includegraphics[width=\textwidth, trim={0 22cm 0 0},clip]{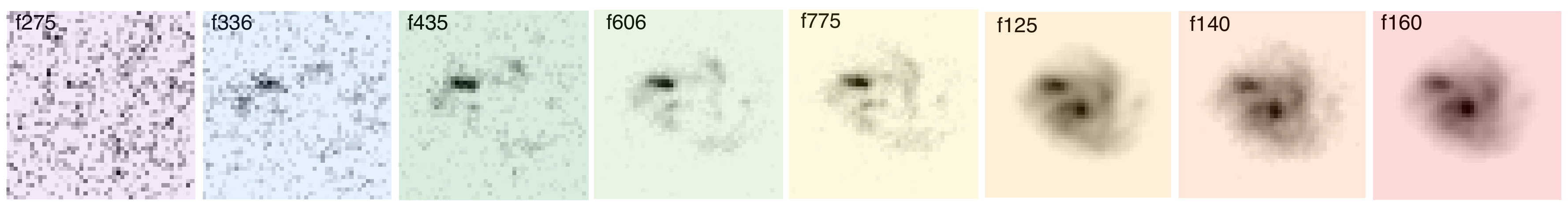}
\caption{U-H band imaging with HST. This galaxy 
exhibits strong color gradients. Going to bluer wave bands, the emission is
increasingly suppressed in the center.
\label{fig:hstims}
}
\end{figure*}

Dust continuum observations were taken with the IRAM NOthern
Extended Millimeter Array (NOEMA) between December 2016 and 
March 2017. 
We observed our primary target, \gn, for a total on-source integration time
of 26 hours in three configurations of the eight antennas:
9h in D, 3h in C and 14h in A (in order from most compact to most extended 
configuration).
Data were taken with the antennas arranged into multiple configurations to efficiently
probe emission on multiple scales. 
We used the compact C and D configurations to probe
faint, extended emission (e.g. from a galaxy disk) and
the extended A configuration to probe bright, compact emission (e.g. from a galaxy bulge).
Observations were carried out in band 3 at 265GHz (1.1\,mm), 
allowing us to measure the \restlam\ dust continuum emission for 
our target at $z=1.248$. 
Although this tuning approached the high frequency/short wavelength limit 
of NOEMA's range where atmospheric transmission is lower, this is 
compensated for by the increasing brightness of the source moving 
up the Raleigh-Jeans tail. 
The conditions varied but were excellent during observations in A configuration. 
The primary source of atmospheric opacity, the percipitable 
 water vapor, was low for the A configuration tracks, $PWV< 1$mm, 
 particularly important for these observations 
 which are at the high frequency extrema of NOEMA's range.
The noise in the system as reflected in the system temperature was $Tsys<200$.
For data taken in the C and D configurations, these values were PWV=3-4mm 
and $200<Tsys<400$. 
Observations of the source were alternated with observations of a bright
quasar every twenty minutes as a calibrator. 
The WideX correlator with a bandwidth of 3.6GHz was used for maximal 
continuum sensitivity.

The data were calibrated following the standard GILDAS/CLIC pipeline performing
absolute flux, bandpass, phase, and amplitude calibrations.
Additional flagging was done by hand. 
We combine the amplitudes and \uv\ distances that comprise the visibility data for all 
configurations of the antennas into a single dataset and
use GILDAS/MAPPING to Fourier Transform the combined data
from \uv\ space to image space using two different weighting schemes. 
We create one image using `Natural' weighting, 
in which each visibility is weighted by the inverse of the noise variance
to maximize point source sensitivity. 
Natural weighting yields an image with a synthesized 
beam size of $0.4$"$\times0.5$" and rms noise of 28 $\mu$Jy/beam.
We create a second, higher resolution image using `Robust' weighting (with threshold=1)
to give increased weight to long baseline data.
This image has a beam size of $0.26$"$\times0.35$" 
and rms noise of 37 $\mu$Jy/beam. 
To improve the resolution in the bright, central region 
where bulge growth is taking place, we replace 
the central resolution element of the natural weighted map with 
the one from the robust weighted map.
We note that in this scheme, flux is not conserved. However, this concern is 
somewhat mitigated because we scale the final image and the surface brightness 
profile to SFR(IR) (as discussed in \S\,\ref{section:sfrindicators}). 
This is our imaging procedure for our default image which is shown in 
Fig. \ref{fig:imsandprofs}.
We use a single clean iteration to correct the absolute flux scale of 
the image but further deconvolution is not warranted by the signal-to-noise
ratio of our data. Instead, when necessary, we convolve our comparison data
sets with to the resolution of the NOEMA data.
\footnote{As a potentially helpful side note:
following the imaging procedure, the interferometric images are in units of Jy/beam. 
To make the flux scale not beam-size-dependent requires scaling
by the beam solid angle for an elliptical gaussian with half power 
beam width major and minor axes $a$ and $b$: $\pi ab/(4{\rm ln}2)$.}

\subsection{Ancillary data: Hubble imaging and spectroscopy + Spitzer \& Herschel photometry}
We leverage our NOEMA data using a wealth of ancillary data, which thanks to 
large investments by the community has been obtained and publicly released.
This includes spatially-resolved imaging from HST in eight bands spanning 
ultra-violet through near-infrared wavelengths as shown in Fig.\ref{fig:hstims}.
The rest-UV is probed by 
F275W, F336W (HDUV, Oesch et al., submitted),
F435W, F606W, F775W \citep[GOODS][]{Giavalisco:04}.
The rest-optical is probed by  F125W, F160W 
\citep[CANDELS][]{Grogin:11,Koekemoer:11},
and F140W \citep[3D-HST]{Skelton:14}.
We use the mosaics provided by the 
3D-HST\footnote{http://3dhst.research.yale.edu/Data.php}
and HDUV\footnote{http://www.astro.yale.edu/hduv/data.html} 
teams \citep[Oesch et al. submitted]{Skelton:14}.
The rest-IR is probed by Spitzer/IRAC 3.6\mic, 4.5\mic \citep{Ashby:13};
5.8\mic, 8\mic\ \citep{Dickinson:03,Ashby:13}; 
and Herschel/PACS 70\mic,100\mic, 160\mic\ \citep[PEP][]{Lutz:11}
We note that these rest-IR tracers are not spatially-resolved 
for galaxies at $z\sim1$.

The redshift of \gn\ was derived based on combined constraints from photometry and 
3D-HST spectroscopy \citep{Brammer:12a, Momcheva:16}.
A stellar mass of $6.8\times10^{10}$\,\msun\ was computed by fitting a stellar 
population synthesis model to the observed U-8\mic\ photometry, 
using \cite{Bruzual:03} templates and 
assuming solar metallicity, a \cite{Chabrier:03} initial mass function, 
an exponentially declining star formation history, and the \cite{Calzetti:00} 
dust attenuation law \citep[see][]{Skelton:14}.

We make an \ha\ map of this galaxy using data from the 3D-HST grism 
spectroscopic survey \citep{vanDokkum:11,Brammer:12a,Momcheva:16}.
To make the \ha\ map of \gn\ we subtract quantitative models for both
the contamination from overlapping spectra of other objects 
and stellar continuum emission. 
The continuum model is generated by convolving the best-fit SED
with the combined $J_{F125W}/JH_{F140W}/H_{F160W}$ image
and accounts for a spatially-resolved stellar absorption. 
With a FWHM spectral resolution of $\sim100\AA$ 
 \ha\,$\lambda6563$\AA\, and [N\,{\sc\,ii}]\,$\lambda\lambda6548+6583$\AA\ 
 are blended.
To account for the contamination of \ha\, by [N\,{\sc\,ii}],
we assume flat radial gradients, scale the measured flux down by a factor of 
\ha$_{\rm corr} =$\ha$_{\rm meas}/1.3$
 and adopt \ha$_{\rm corr}$ as the \ha\ flux \citep{EWuyts:14,EWuyts:16}. 
For a more detailed description, see \cite{Nelson:16b}. 
We use iraf PSFMATCH to convolve all the HST data to the 
resolution of the millimeter data.

\begin{figure}
\centering
\includegraphics[width=0.5\textwidth]{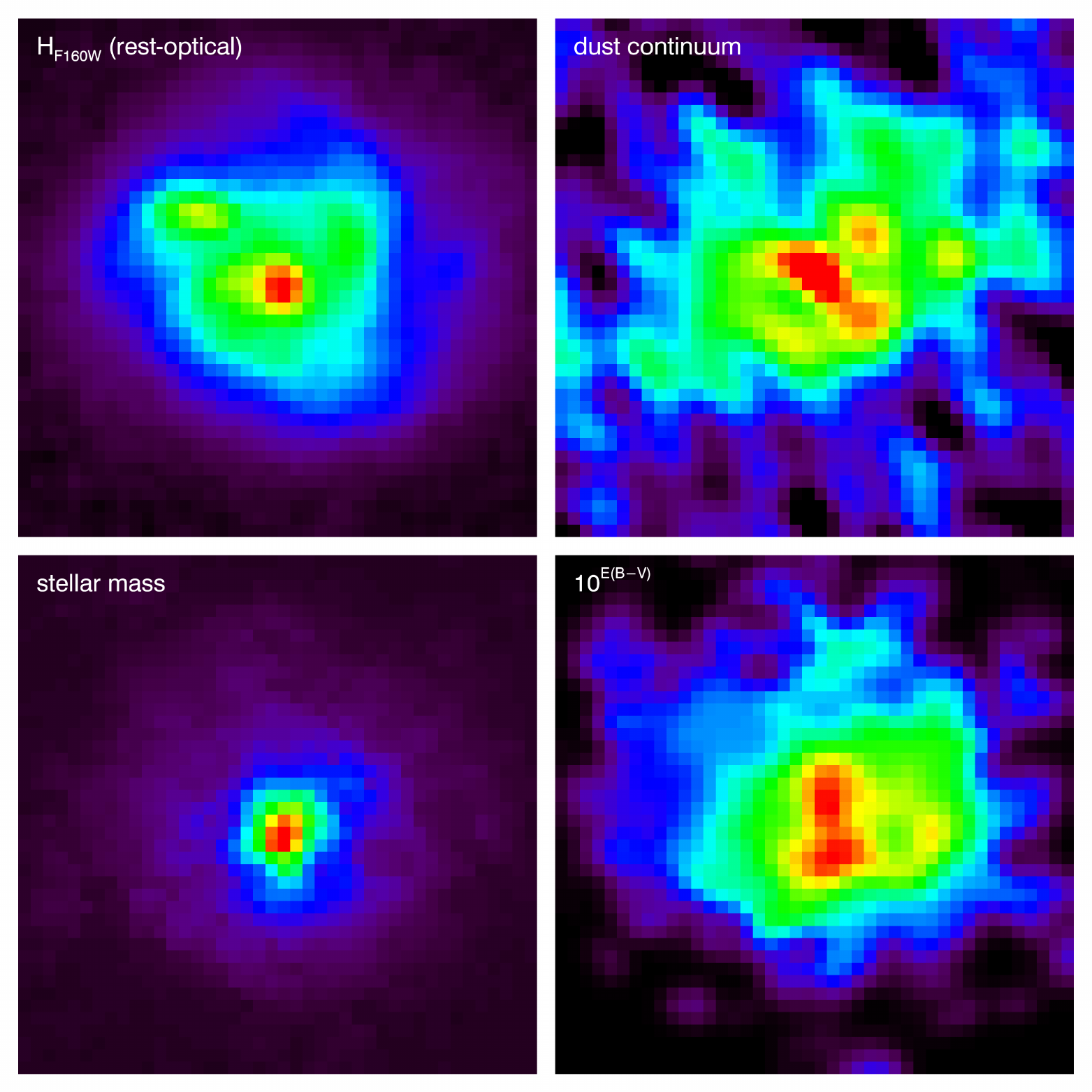}
\caption{The top row shows observed quantities: the HST \h\ and
NOEMA dust continuum images. The bottom row shows maps derived
from spatially-resolved stellar population synthesis modeling: 
maps of stellar mass and dust reddening. 
\label{fig:sps}
}
\end{figure}

\subsection{Spatially-resolved stellar population properties}
Spatially resolved maps of stellar mass and E(B-V) made from the 
eight band HST imaging are shown in Fig.\,\ref{fig:sps}.
The creation of these maps is covered in detail in \cite{Cibinel:15}
but we briefly describe it here for completeness. 
Image postage stamps are cut from the mosaics in each HST band
(Fig. \ref{fig:hstims}). 
Already from the images in different bands it is clear that this galaxy 
exhibits strong color gradients: redder in the center and bluer at larger radii. 
The reddening can be due to age, dust, or both and will affect inferences about 
the distribution of star formation and stellar mass.
Rest-UV colors can be used to help distinguish between 
the effects of dust and age \citep[see e.g. ][]{Liu:17}.
Postage stamps are convolved by psf-matching to the resolution of the 
reddest band (\h, which has the lowest resolution). 
These images are then adaptively smoothed using Adaptsmooth
\citep{Zibetti:09} requiring $S/N>5$ in each spatial bin
 in the \h\ image, which has the highest $S/N$. 
A potential cause for concern is that this may result in an additional 
smoothing in the galaxy center where the surface brightness is
changing most rapidly. 
This concern is alleviated, however,  by noting that the center is bright and has 
high $S/N$ in \h\ so no smoothing is done inside 6\,kpc.
The SPS code LePhare \citep{Arnouts:99,Ilbert:06}
is run on the photometry in each spatial bin using the \cite{Bruzual:03}
synthetic spectral library, a \cite{Chabrier:03} IMF, 
a \cite{Calzetti:00} dust law with $0<E(B-V)<0.9$mag, 
and three metallicity values ($Z=0.2,0.4,1\,Z_\odot$).
We adopt a delayed exponential star formation history 
$(t/\tau^2)\,{\rm exp}\,(-t/\tau)$ with a characteristic timescale $\tau$ 
with 22 values between 0.01 and 10 Gyr and a minimum age of 100\,Myr. 
This method is qualitatively similar to that described in \cite{Wuyts:12} 
and \cite{Lang:14}. Despite using slightly different assumptions and 
different SPS codes, within 8\,kpc, the stellar mass maps from these
two methods are typically the same to within 20\%.

\subsection{Star formation indicators}\label{section:sfrindicators}
Fig.\,\ref{fig:imsandprofs} shows the three different tracers of star formation
we have for \gn: UV, \ha\ and submm. 
First, the rest-UV (1216-3000\AA) traces emission from stars with lifetimes 
<100Myr and the following can be used to scale the UV luminosity to 
a star formation rate on this timescale:
$$SFR(UV)=2.40\times10^{-10}L_{UV} [L_\odot]$$
$L(UV)$ is the total UV luminosity from $1216-3000$\AA\ computed by 
scaling the rest-frame 2800\AA\ luminosity \citep{Bell:05,Whitaker:14}.
We probe the rest-UV emission with the HST/$B_{F435W}$ filter, 
corresponding to rest-frame 2040\AA, near the center of the optimal 
wavelength range for determining UV-based star 
formation rates. We then scale this image to the integrated UV SFR.  
Because of the short wavelength, this emission is 
highly attenuated by dust.

Second, the \ha\ recombination line re-emits emission shortward of the Lyman limit,
providing a probe of stars with lifetimes <10Myr. 
To scale the \ha\ luminosity to a star formation rate, we use the 
relation presented  in \cite{Kennicutt:98araa} adapted from a 
Salpeter to a \cite{Chabrier:03} IMF: 
$$SFR(H\alpha)=1.7\times10^{-8} L_{H\alpha} [L_\odot]$$
With a rest wavelength of 6563\AA, \ha\ is less impacted by 
dust attenuation than the UV but still suffers significant attenuation
in dusty galaxies. 

Third, the bulk of the bolometric luminosity from young massive stars is absorbed 
and re-radiated in the infrared, with a peak in emission near 100\mic\ (e.g. Lutz et al. 2016). 
The total infrared luminosity ($L(IR)$) is computed from Herschel/PACS
160\mic\ \citep{Lutz:11,Magnelli:13} 
using a luminosity-independent template \citep{Wuyts:11a} and 
scaled to a star formation rate using: 
$$SFR(IR)=1.09\times10^{-10}L_{IR} [L_\odot]$$
In intermediate- and high- redshift galaxies, this far-IR emission is not
resolved with Herschel.  The only currently feasible way
to resolve long-wavelength dust emission in intermediate- 
and high-redshift galaxies is using mm/submm interferometers. 
Here we take advantage of the high resolution and continuum sensitivity of 
the NOEMA millimeter interferometer to resolve the \restlam\ dust continuum
emission. 
We scale this \restlam\ image to the total SFR(IR) computed from the 
Herschel/PACS 160\mic\ and use it as a proxy for dust-obscured star formation.
In star-forming galaxies, emission at the IR peak is due largely to thermal 
emission by dust that has been heated by star formation. 
Our scaling assumes a flat temperature gradient across 
the galaxy, while in local galaxies typically exhibit negative temperature 
gradients (i.e. hotter temperatures in the center) 
\citep[e.g.][]{Engelbracht:10,Pohlen:10, MGalametz:12,Hunt:15}. 
If this were the case in this galaxy, it would suggest that the intrinsic 
dust-obscured star formation is more centrally concentrated than we infer.


\begin{figure*}
\centering
\includegraphics[width=\textwidth]{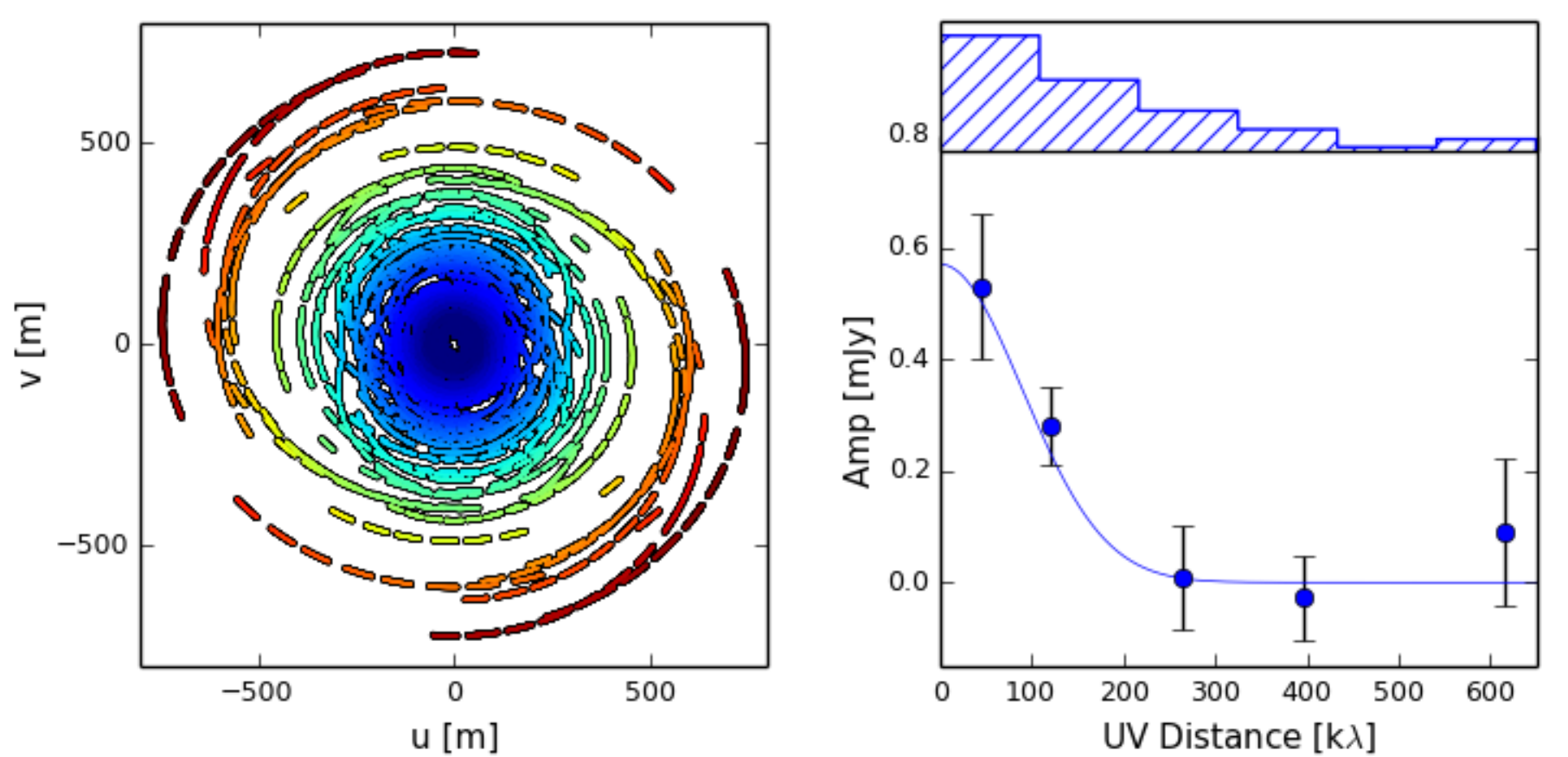}
\caption{We fit the size of the \restlam\ continuum in the \uv\ plane. Left: data 
as a function of position in the \uv\ plane. Right: Data averaged in bins of UV distance,
with the fit overplotted as a line. The histogram reflects the quantity of data in 
each bin of \uv\ distance 
\label{fig:mmsizefit}
}
\end{figure*}

\section{Structural properties of galaxy growth}
Fig. \ref{fig:imsandprofs} shows the images and radial surface brightness 
profiles of the different tracers of star formation in \gn. What is immediately 
clear is that the dust-obscured star formation is more concentrated 
than the unobscured star formation. 
In this section we quantify the structural properties of growth in \gn\ using the 
size, concentration, and radial surface brightness profiles. 
We compare these quantities amongst the different star formation tracers
UV, \ha, and dust continuum emission and between stellar mass and 
rest optical light. By comparing the structural parameters of star formation
to those of the stellar mass, we infer how this galaxy is growing through
star formation.

\subsection{Size}
We measure the size of the dust continuum emission directly from the visibility data
(as shown in Fig.\,\ref{fig:mmsizefit}).
The advantage of fitting the observed visibilities directly in the interferometric 
\uv-plane, rather than in the image plane is that the uncertainties associated
with the complex mathematical transformations performed in the imaging process
are removed.  
In this method, each model flux distribution is convolved with the beam in 
image space, Fourier Transformed to \uv-space, then resampled to the observed
UV baselines. 
The fitting is performed using a circular gaussian model with the centroid, flux, and 
full width at half maximum (FWHM) as free parameters.
The best fit is then determined by $\chi^2$ minimization.
We find $r_e($\restlam$)=3.4\pm 0.7$\,kpc. 
We obtain this fit using the GILDAS/Mapping routine UVFIT, results are the same 
within the errors when using CASA/UVMODELFIT. 
If we instead fit with the physically more well-motivated exponential, we find 
$r_e($\restlam$)=3.7$\,kpc, well within the uncertainties of the fit.

We measure sizes for the rest-optical data 
(specifically $JH_{F140W}$ tracing the $\lambda_{rest}=6220$\AA light)
 and stellar mass map by fitting S\'ersic models \citep{Sersic:68}
convolved with the point spread function (PSF) of the images 
using GALFIT \citep{Peng:02}.
For the mass map, we used the empirical \h\ PSF (the PSF to which 
all images that go into making the mass map are convolved).
For the $JH_{F140W}$ image we use the interlaced PSF generated by Tiny Tim
\citep{Krist:95}.
We determine error bars by performing Monte Carlo simulations forcing the values of the 
centroid and S\'ersic index to remain fixed. We vary the centroid within
a 0.2" box and the S\'ersic index by $\pm50\%$. 
Neither the \ha\ nor UV emission are centrally peaked so they are poorly fit
by S\'ersic models and correcting for the PSF is unimportant for a determination
of the size. Instead, we measure their sizes using growth curves.
The radial surface brightness profiles are measured in finely sampled circular 
apertures and $r_e$ is the radius at which the the enclosed flux
is 50\% of the total.
We determine the uncertainties on the growth curve sizes by taking the standard 
deviation of Monte Carlo simulations run varying the centroid within a box of 0.2".
We find $r_e($optical$)=4.7\pm0.2$\,kpc, $r_e($mass$)=2.9\pm0.2$\,kpc,
$r_e(UV)=6.2\pm0.2$\,kpc, and $r_e($\ha$)=4.8\pm0.2$\,kpc, 

With regard to star formation tracers, we find $r_e(UV) > r_e($\ha$) > r_e($submm).
That is, the submm, which traces dust-obscured star formation, is the most compact,
the UV, which traces unobscured star formation is most extended, and the \ha\ which
is somewhat less impacted by dust attenuation is in the middle. 
Given that the UV size is nearly twice that of the submm size, it 
is clear that gradients in dust attenuation play a significant role in this galaxy. 
The true distribution of star formation is more compact than would be inferred 
based on the \ha\ or UV emission alone. 
If \gn\ has a negative temperature gradient, then the dust-obscured star formation
would be even more compact than we measure.

We find that the stellar mass distribution inferred from spatially-resolved SED fitting
is more compact than the rest-optical light. 
Comparing the distribution of star formation to existing stellar mass, we find
$r_e(UV) > r_e($\ha$) \geq r_e($optical$) > r_e($submm$) \geq r_e($mass).
Thus, the dust emission is similar to or more extended than the stellar mass. 
Even though the the star formation is more compact than the unobscured 
tracers suggest, the mass is also more compact than the rest-optical light 
suggests. 
This galaxy is still building between self-similarly and inside-out (i.e. growing larger in size due
to star formation).

\begin{figure}
\centering
\includegraphics[width=0.5\textwidth]{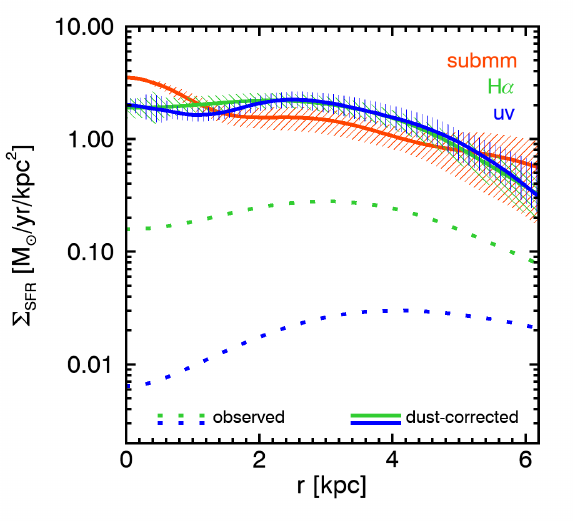}
\caption{Here we show the radial distribution of star formation inferred from the 
unobscured tracer \ha\ alone (light blue), corrected for dust attenuation using the
rest UV-optical SED (blue), and combining the obscured plus unobscured 
tracers \restlam\ and \ha\ (red) as described in \S \ref{section:sfrprofs}. 
After correcting for dust using the SED Av map, the \ha\ traces the IR 
fairly well in a radially averaged sense. There does appear to be some excess
central emission, however that is too obscured to recover. 
\label{fig:dustcorr}
}
\end{figure}

\subsection{Concentration} \label{section:concentration}
We compute the concentration by dividing the flux in a central aperture by the total flux,
As an estimate of concentration we measure $C=F(r<0.3")/F(r<1")$,
a number which is related to the bulge to total ratio in a galaxy 
\citep[e.g.][]{Abraham:94,Abraham:96,Lotz:04}.
We choose this definition of concentration to optimally use the information 
content of the new interferometric millimeter data presented in this paper. 
We measure $F(r<0.3")$ from our higher resolution, lower sensitivity image
and $F(r<1")$ from our lower resolution, higher sensitivity image. 
We find a concentration of $C($\restlam$)=F(r<0.3")/F(r<1")=0.17$.
We can compare this concentration to the concentration of other star formation tracers
\ha\ and UV as well as to optical emission and stellar mass.
We measure concentrations in an identical way on the convolved HST data
and find $C($\ha$)=0.04$, $C({\rm UV})=0.02$, $C({\rm optical})=0.10$, 
$C({\rm mass})=0.22$.

Analogously to the trends for effective radius we find 
$C($UV$) < C($\ha$) < C($optical$) < C($\restlam$) <  C($mass).
The millimeter emission is more concentrated than the \ha\ emission which is
more concentrated than the UV emission. The dust obscured star formation
is more centrally concentrated than the unobscured star formation: 
the dust-obscured star formation is growing the bulge. 
The \restlam\ emission is also more centrally concentrated than the
rest-optical emission but not than the stellar mass. 
Based on comparing the size and concentration of the \restlam\ emission to the 
the rest-optical, which is often taken as a proxy for stellar mass, 
we would infer that the star formation is more compact
than the stellar mass implying that star formation is actually shrinking 
the effective radius of the galaxy. 
However, if we instead compare the size and concentration of the \restlam\ emission 
to those of the modeled stellar mass map, this is not the case. 
The bulge of this galaxy is undergoing a period of growth but this growth
is not so dramatic that it decreases the effective radius of the galaxy. 

\subsection{Radial profiles of SFR and sSFR } \label{section:sfrprofs}

\begin{figure}
\centering
\includegraphics[width=0.5\textwidth]{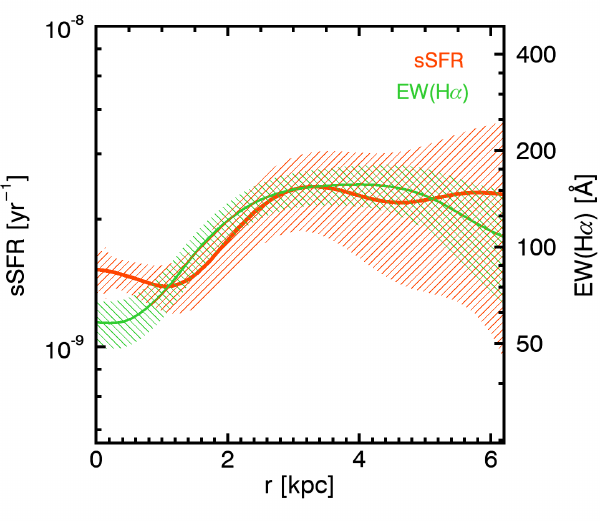}
\caption{Here we show a comparison of the radial \ha\ equivalent (EW(\ha)) 
and specific star formation rate (sSFR) and profiles for \gn. 
In blue, EW(\ha) reflects the quotient of the \ha\ and surrounding continuum 
emission, which is often taken as a proxy for sSFR. 
In red, the plotted sSFR is the quotient of the SFR traced by the \restlam\
continuum emission and the stellar mass. 
The sSFR is somewhat higher at large 
radii than in the center, suggesting the stellar mass is growing
more rapidly in the outskirts; the galaxy is building inside-out.  
The radial behavior of EW(\ha) is fairly similar to the dust-corrected sSFR in this galaxy,
meaning that in this galaxy EW(\ha) is a good tracer of the sSFR.  
\label{fig:ssfr}
}
\end{figure}

Finally we consider the radial SFR profiles in \gn,
comparing the obscured and unobscured tracers of star formation
(Fig.\,\ref{fig:dustcorr}). 
We additionally show where this galaxy is growing by comparing the radial distribution
of star formation and stellar mass using the specific star formation rate
$sSFR=SFR/M_*$ (Fig.\,\ref{fig:ssfr}). 
We extract radial profiles from the PSF-matched images of star formation,
made as described in \S\,\ref{section:sfrindicators}. 
This galaxy is fairly round ($b/a=0.83$ in \h) so we extract radial profiles in circular 
apertures centered on the \h\ flux-weighted centroid (which is also the center of mass).

The radial profiles of star formation are shown in Fig.\,\ref{fig:dustcorr}
(as well as Fig.\,\ref{fig:imsandprofs}).
The dust-obscured star formation dominates over the unobscured star
formation at all radii. 
It is roughly an order of magnitude greater than the \ha\ based star
formation rates and nearly two orders of magnitude greater than the 
UV-based star formation rates. 
But it is not just an offset between the three tracers, the profiles 
also have markedly different shapes, reflecting the results from simpler
size and concentration measurements. 
While the \ha, and to an even greater extent the UV, are centrally depressed, 
the dust is centrally peaked. 
There is significantly more dust-obscured
bulge growth than implied by the unobscured tracers.

Clearly, without accounting for dust, the radial distributions of 
star formation inferred from different tracers are very different. 
We test using a map of dust attenuation from spatially-resolved SED
modeling to correct the \ha\ and UV emission for the effects of dust. 
(Note, the SED modeling includes four bands covering rest-NUV-FUV: 
ACS/F606W \& F435W plus UVIS/F336W \& F275W.)
We use the resulting E(B-V) map 
to correct the SFR(\ha) and SFR(UV) for the effects of dust with: 
$$SFR(dustcorr)=SFR\times10^{0.4k(\lambda)E(B-V)}$$  
where $k(\lambda)$ is computed using the \cite{Calzetti:00} 
dust attenuation law at the wavelength of our observations
 $k(H\alpha=6563$\AA$)=3.32$ and $k(UV=2040$\AA$)=8.76$.
 As our default, we do not include extra attenuation toward HII regions
 when correcting the \ha\ but the difference in the profile when adding
 extra attenuation based on \cite{Wuyts:13} is included the error bar.
 Error bars are additionally comprised of the formal error on the 
 SPS fit of the dust attenuation and the noise in the image. 
The overall star formation scale is set by the integrated
SFR(IR) so this is a test of the radial dust attenuation estimate
not the overall scaling.

The dust-corrected radial profiles are shown in Fig.\,\ref{fig:dustcorr}. 
There is surprisingly good agreement between IR and the dust-corrected \ha\ and UV,
especially considering how dramatic the differences were before the dust correction.
The exception is the center where there is somewhat more dust-obscured star formation
than inferred by the dust-corrected \ha\ and UV.
This suggests that the dust geometry may be more complex 
than the simple foreground screen assumed for the spatially-resolved SED fitting: 
if the stars and dust are mixed and some regions have $\tau>>1$,
the rest-optical/UV colors may fail to recover the total quantity of dust-obscured 
star formation.
However, the differences are small and the primary 
take-away from this exercise is that the dust attenuation map from 
spatially-resolved SED-fitting 
to the UV-NIR imaging performs reasonably well correcting for dust 
in this galaxy.

Fig. \ref{fig:ssfr} shows the radial profile of the specific star formation rate, 
the star formation rate per unit stellar mass ($sSFR = SFR/M_*$), 
a reflection of the rate of growth relative to the stellar mass already present.
This quantity is derived as the quotient of the star formation rate surface density
inferred from the dust continuum emission and the stellar mass map 
convolved to the same resolution.
Also shown in Fig. \ref{fig:ssfr} is the radial \ha\ equivalent width profile,
effectively the scaled quotient of the \ha\ and respective broad-band emission,  
which is often used as a tracer of sSFR.
Comparing the radial profiles of EW(\ha) and the dust continuum based sSFR,   
we find that inside $r<5$\,kpc, the two are similar. 
This similarity suggests that in this galaxy, EW(\ha) is an effective
tracer of sSFR and can be used to determine where a galaxy is growing.
We find that the sSFR increases radially meaning that the galaxy is growing 
faster in the outskirts than in the center. 
This positive sSFR gradient is consistent with the idea that star formation, even after 
correcting for dust attenuation is more extended than the 
existing stellar mass. The star formation is making the galaxy larger, 
growing it from the inside-out. 

\section{Discussion}

\begin{figure}
\centering
\includegraphics[width=0.5\textwidth]{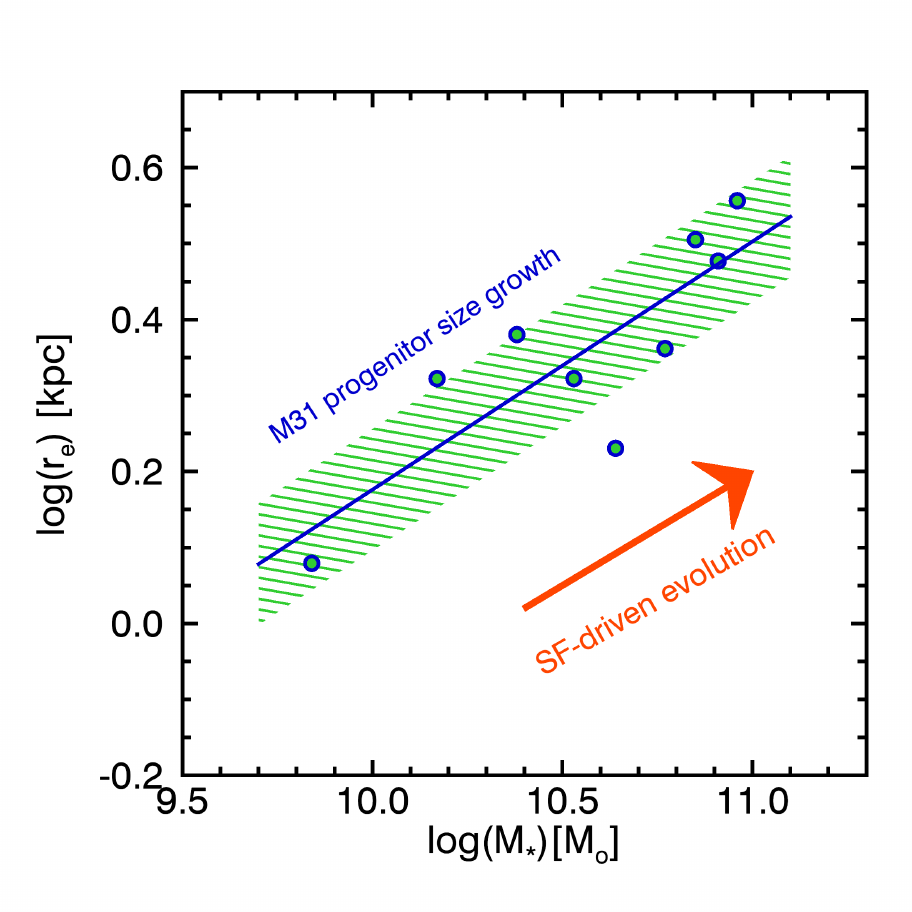}
\caption{Size growth implied by dust-obscured star formation (orange arrow)
relative to that empirically derived for M31 progenitors by abundance matching. 
The points are from \cite{Papovich:15}, the blue line shows the best fit,
and the green hatched region shows the $1\sigma$ deviation of the 
points from the fit. The size growth implied by the dust-obscured star formation 
is consistent with the expected size growth from population studies. 
\label{fig:sizegrowth}
}
\end{figure}

\subsection{Evolution in Size} \label{section:sizeev}
To understand how the observed star formation contributes to structural evolution,
we determine to what extent the star formation can account for the structural
evolution we expect based on known population-wide scaling relations. 
First we compare the size growth of \gn\ due to star formation to the size growth of 
Andromeda progenitors empirically derived by \cite{Papovich:15} using abundance matching. 
With $SFR=169$\,\msun/yr, in a time of $\Delta t= 100$\,Myr,
 this galaxy will grow by \deltam$=1.7\times10^{10}$\msun\ 
(to $8.5\times10^{10}$\msun). 
To estimate the size evolution due to this star formation, we scale the best-fit
model for the \restlam\ emission to \deltam\ and add it to the psf-corrected
profile of the existing stellar mass.
Thus, we sum the radial profiles \m(r) and \deltam(r) and measure the effective radius 
of the resulting radial profile using a growth curve. 
We do not constrain the profile shape so as a check we also do this exercise 
with S{\'e}rsic indices $n=0.3-3$.

\begin{figure}
\centering
\includegraphics[width=0.5\textwidth]{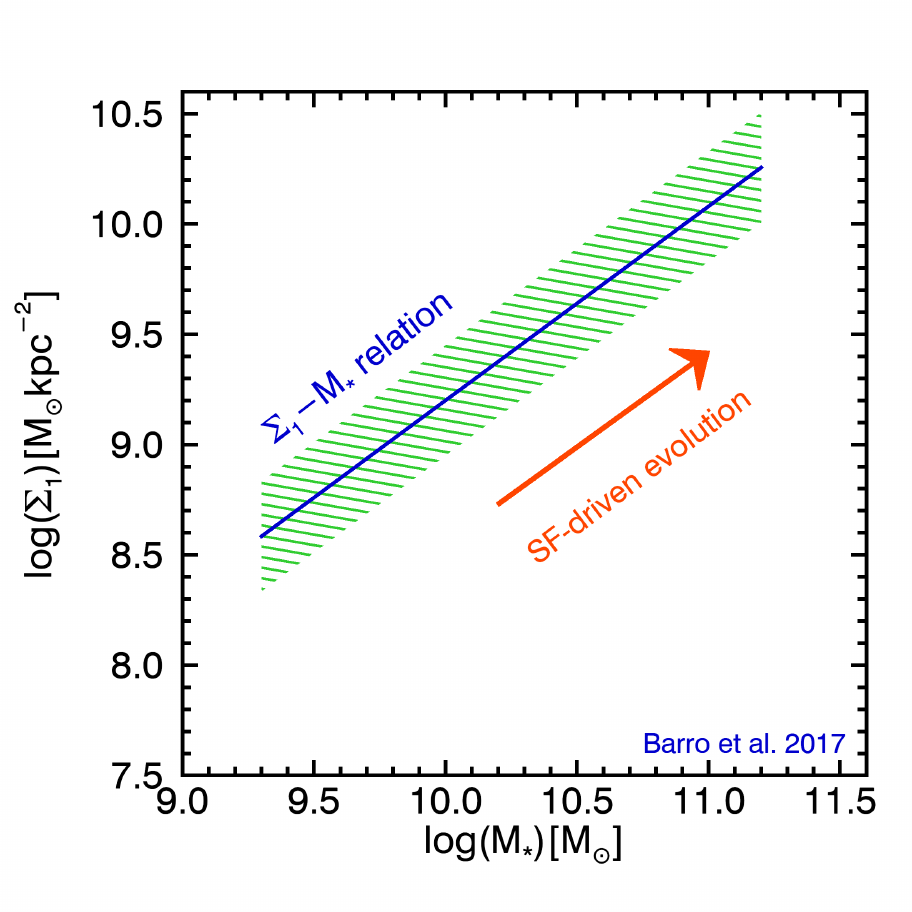}
\caption{Build-up of central stellar mass density implied by dust-obscured 
star formation (orange arrow) relative to the \sigmaone-\m\ relation 
for star-forming galaxies of
\cite{Barro:17} (blue line). The green hatched area shows the 
$\sigma({\rm log}\Sigma_1)\sim0.25{\rm dex}$ observed scatter in the relation. 
The starting point of the arrow is offset arbitrarily because we measure 
$\Sigma(r<2 {\rm kpc})$ rather than the $\Sigma(r<1 {\rm kpc})$ that 
\citep{Barro:17} measure. 
\label{fig:sigma1}
}
\end{figure}

Our inferred growth trajectory based on dust-obscured star formation is shown 
as the orange arrow in Fig.\,\ref{fig:sizegrowth}, while 
the size evolution of the population of Andromeda progenitors is shown by the
points and blue line. 
The size growth inferred due to star formation is 
$\Delta r_e/\Delta M_* = 0.3\pm0.1 $.
The uncertainty includes results using S{\'e}rsic indices $n=0.3-3$.
The average size growth of the population of the Andromeda progenitors based 
on a linear fit to the \cite{Papovich:15} size measurements is 
$\Delta r_e/\Delta M_* \sim0.3 $, a growth rate consistent with star forming
galaxies in general \citep{vandokkum:15}.
In this galaxy at this time, the expected size growth trajectory can in principle 
be explained simply by the addition of stellar mass due to star formation
(without the need to invoke other processes like merging to redistribute angular 
momentum after the stars are formed). 
However, the uncertainty on our \restlam\ size measurement is still 
large and including the full range of possibilities in this analysis means
that $\Delta r_e/\Delta M_*$ can formally range from slightly negative to
$\sim0.5$, so we caution against interpreting this aspect of our analysis too 
strongly.

\subsection{Evolution in Central Surface Density} \label{section:csdev}
We also investigate the evolution of the central stellar mass density of this
galaxy, due to dust-obscured star formation. 
Specifically, we determine to what extent the growth of the central stellar mass
density can be accounted for by star formation. 
To do this, we use the central stellar mass surface density
$$\Sigma_{cen}=M_{cen}/r_{cen}^2$$ 
where $M_{cen}$ is the mass contained in a central aperture with 
radius $r_{cen}$.
There is a relation between this central stellar mass surface density 
and the total stellar mass of galaxies \sigmacen-\m\ 
\citep{Barro:17}.
This relation results from the combined effect of the $M_*-r_e$ and $M_*-n_{sersic}$\, 
relations and encapsulates the trend of increasing bulge dominance with increasing stellar mass.
The key question is: is the star formation we observe
consistent with bulge building, moving the galaxy along this relation?

To answer this question, we compute the trajectory of \gn\ in the 
\sigmacen-\m\ plane due to star formation. If star formation can 
account for bulge growth, this trajectory should move the galaxy 
along the observed \sigmacen-\m\ relation. 
Investigating this question requires at least a first order correction 
for the PSF/beam the images. To estimate a correction for the PSF/beam
we use S\'ersic models as described in \cite{Szomoru:10}. 
Briefly, we use GALFIT best-fit parameters to produce a model for
the galaxy that is not convolved with the PSF/beam. 
We then add the S\'ersic model residuals back to this 
unconvolved model to produce an image that has a first order correction 
for the PSF.
The fits for the HST data are stable and trace the data well. 
The fit for the NOEMA data is not stable with different initial conditions producing different fits. 
With the instability of this fit, we tested a large range of fit parameters and 
found that as long as the central aperture we used was larger than the beam, 
the measurement was robust against varied fitting parameters. 
Hence, rather than the 1\,kpc aperture used by \cite[\sigmaone,][]{Barro:17}, we use a 
2\,kpc aperture and call this value \sigmacen.
Our error bar on this measurement includes the full range of beam 
corrections derived in the fitting.

In Fig.\,\ref{fig:sigma1}, we show a comparison between the 
population \sigmaone-\m\ relation found by \cite{Barro:17} (blue line)
and the trajectory \gn\ is moving in this plane due to the star formation we observe
(orange arrow).
We find that the star formation building this galaxy is moving it along
the \sigmaone-\m\ relation, suggesting that as star formation adds mass
to this galaxy, we are witnessing the growth of its bulge. 
To the best of our information in this galaxy, star formation can build a bulge 
consistent with the structural relation of star forming galaxies at this epoch.

\begin{figure}
\includegraphics[width=0.5\textwidth]{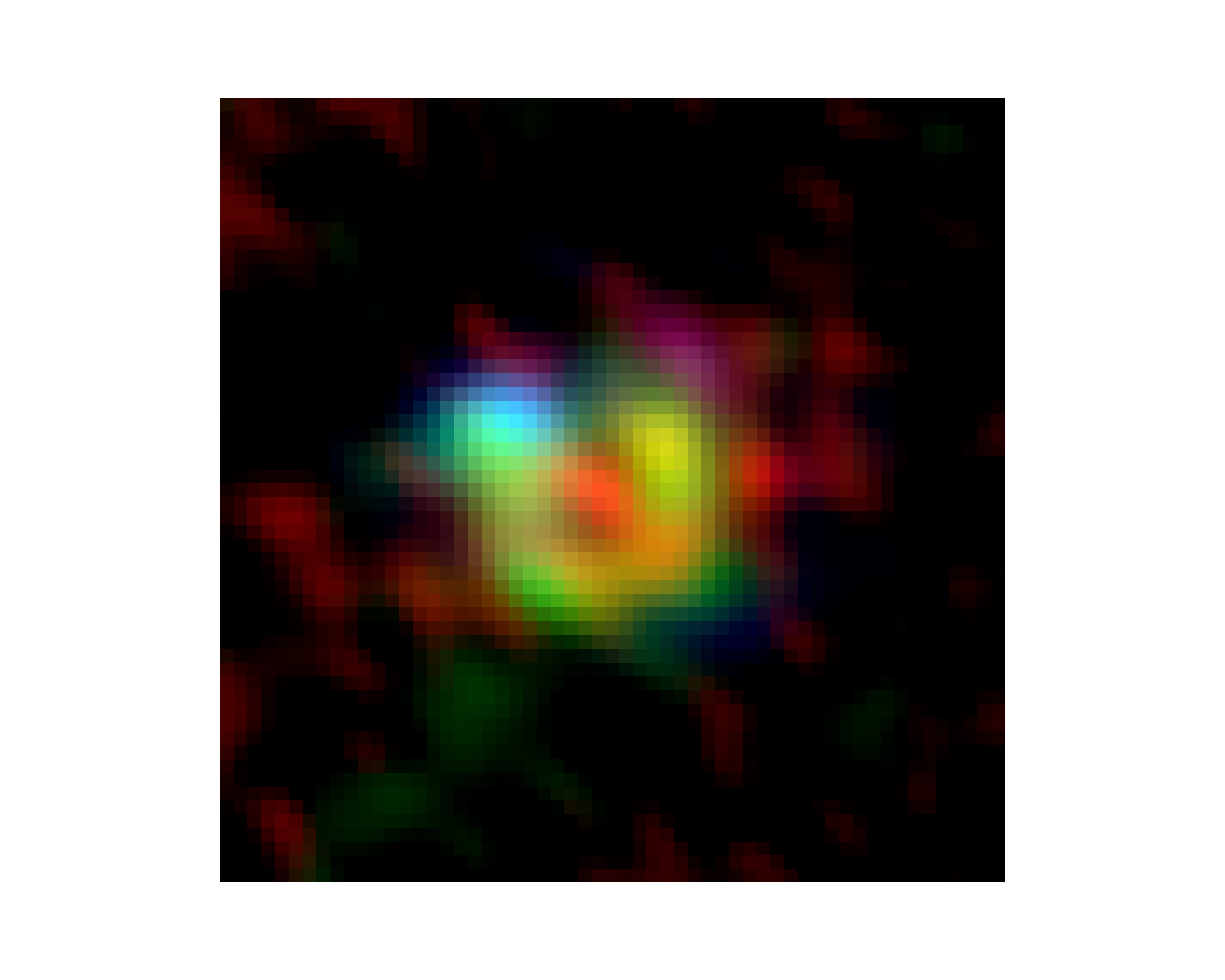}
\caption{Three color image of star formation with UV in blue, \ha\ in green, and \restlam\ in red. 
The three different tracers trace distinctly different regions of the galaxy.
\label{fig:colorim}
}
\end{figure}

\begin{figure*}
\centering
\includegraphics[width=0.7\textwidth]{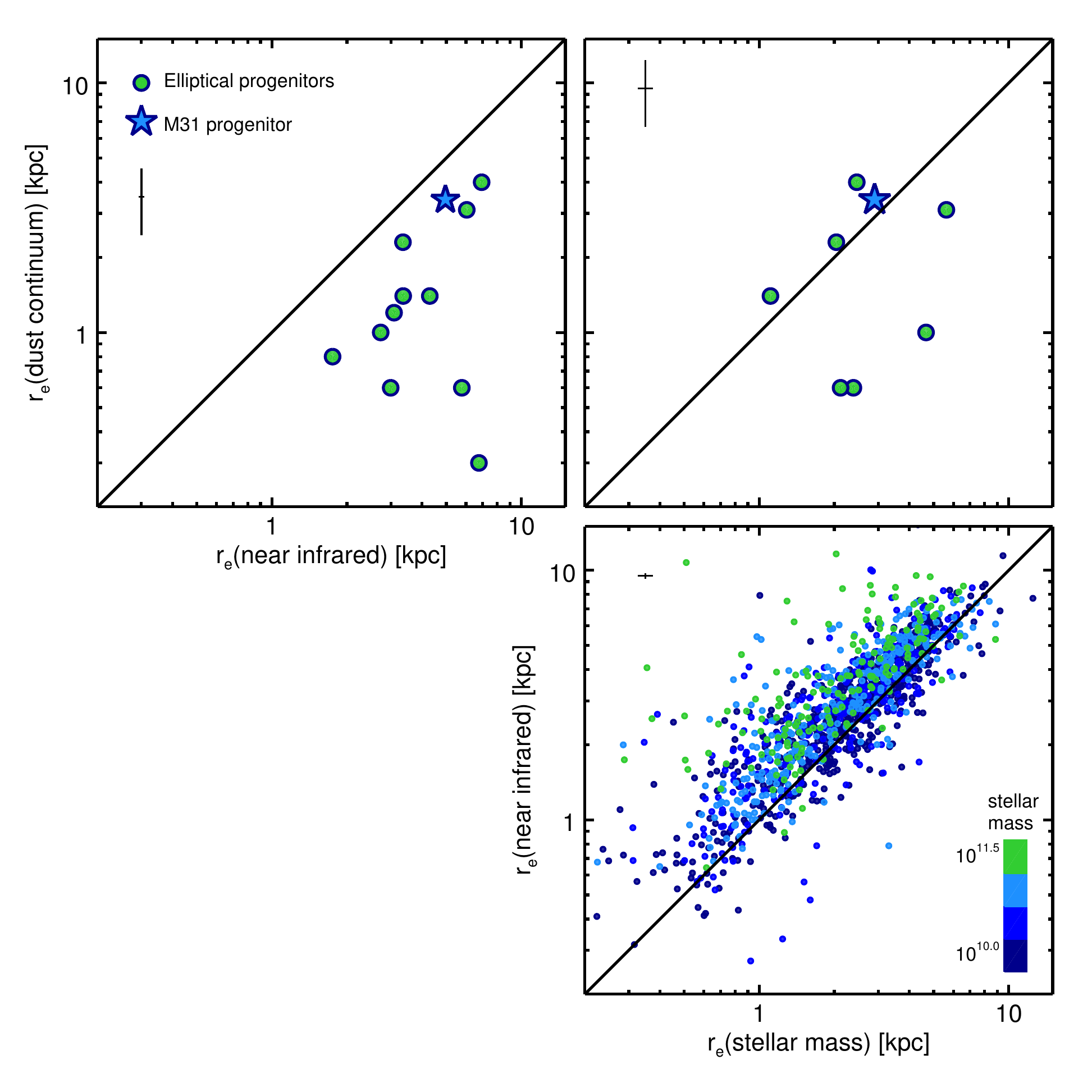}
\caption{Size comparison of the M31 progenitor \gn\ 
($M_*=6.76\times10^{10}{\rm M_\odot}$ at z=1.25) and the 
massive elliptical galaxy progenitors of \cite{Tadaki:17a}
($M_*>1\times10^{11}{\rm M_\odot}$ at z=[2.19,2.53]).
The left panel shows that in all the galaxies, the dust 
continuum is more compact than the near infrared (rest-optical) continuum 
emission. The bottom right panel shows that in most galaxies
with $M>1.6\times10^{10}$ at $1<z<2.5$, the stellar mass is more 
compact than the near infrared continuum emission. 
The top right panel shows that in roughly half the galaxies galaxies in these studies
the submm is more compact than the stellar mass and in half it is more extended. 
Taken at face value, this would suggest that half the galaxies are growing
more compact due to star formation and half are staying the same size or 
growing slightly larger. 
However, the very high central dust column densities may result in the central
stellar mass densities being underestimated. This would mean that the stellar
mass sizes are in fact smaller than plotted here, moving the points to the left.   
Note that four galaxies are not present here because we did not conduct 
spatially-resolved mass modeling for galaxies with $H_{F160W}>23.5$
\citep[see][]{Wuyts:12,Lang:14}. 
The crosses in the upper left corner of each plot show the measurement 
uncertainty. For the near infrared and dust continuum radii, this is the average
uncertainty from \cite{vanderWel:14} and \cite{Tadaki:17a} respectively. 
For the stellar mass, this is the fit uncertainty measured for \gn.
However, with high dust column densities, the systematic uncertainty on the 
stellar mass sizes is larger. 
\label{fig:ellipcomp}
}
\end{figure*}

\subsection{Comparison to elliptical progenitors} \label{section:ellipticalcomp}
In addition to placing the dust-obscured star formation in \gn\ in the context 
of structural growth, we also compare it to other dust continuum size 
measurements at intermediate redshift. 
In particular we consider the growth patterns in this \prog\ at $z\sim1$ 
to those in massive elliptical progenitors at $z\sim2$.
We compare to \cite{Tadaki:17a} who present size 
measurements for \ha-selected galaxies with 
$M_*>10^{11}$ at $z=2.19$ and $z=2.53$.
No lower mass galaxies were detected with sufficient fidelity to measure a size 
hence there are no size measurements for galaxies with $M_*<10^{11}$.
\cite{Barro:16b} also studied dust continuum sizes in massive galaxies
at $z\sim2$. However, we do not include them here as they 
were specifically selected to be compact in optical light, which complicates
the comparison.

First we compare rest-optical and dust continuum sizes
To probe similar rest-optical wavelengths we use \h\ sizes for the massive 
comparison sample of galaxies at $z\sim2$ and the \j\ size for \gn\ at $z=1.25$, 
corresponding to rest wavelengths
of $\sim5000$\AA\ and 5900\AA\ respectively \citep[from CANDELS]{vanderWel:14}.
As shown in the left panel of Fig.\,\ref{fig:ellipcomp}, in all galaxies, 
the dust continuum is more compact than the rest-optical.
This means that all galaxies studied here display significant dust gradients,
and likely dust attenuation gradients, 
although the latter depends on the geometry of the dust.
In \gn\ the submm size is more compact than the optical size by a factor of 
$r_e(optical)/r_e(submm)=1.4$; 
amongst the massive, $z\sim2$ comparison sample, 
it is on average a factor of 2 ($r_e(optical)/r_e(submm)=2.0$). 
Relative to the rest-optical, the dust continuum in \gn\ is less dramatically compact than 
in the massive elliptical progenitors; it is similar to the galaxies 
of \cite{Tadaki:17a} with the most extended dust continuum emission. 

Second, we compare the rest-optical and stellar mass sizes. 
While the dust continuum is nearly always more compact than the rest-optical 
emission, if these galaxies have color gradients, they will also have $M/L$ gradients.
Fig.\,\ref{fig:ellipcomp} shows $r_e(M_*)$ vs. $r_e(light)$ for the full population of 
galaxies as computed in \cite{Lang:14}.
This shows that the effective radius is almost always smaller in mass than in light: 
the stellar mass is nearly always more compact than the rest-optical light.

Finally, we compare the dust continuum and stellar mass sizes. 
The top right panel of Fig.\,\ref{fig:ellipcomp} shows 
$r_e(submm)$ vs. $r_e(M_*)$ for all galaxies for 
which we have a measurement of the stellar mass effective radius. 
\gn\ has $r_e(submm)>r_e(M_*)$ as do 3/7 galaxies in the sample of \cite{Tadaki:17a},
while the remaining 4/7 of galaxies in this sample have $re(submm)<r_e(M_*)$.
At face value this means there 
is a mix of dust continuum sizes, both larger and smaller than the stellar mass sizes. 
Taking the sub-millimeter emission as a proxy for star formation, 
$r_e(submm)<r_e(M_*)$ could be explained by galaxies undergoing a 
compaction event in which their dense central regions are grown by 
a dissipative event which brings gas to the center inducing central star formation
\citep[e.g.][]{Dekel:14,Zolotov:15}. 
On the other hand, $r_e(submm)>r_e(M_*)$ most simply implies 
that at the time of observation, the galaxy is building inside out,
with star formation increasing the effective radius
\citep[e.g.][]{Nelson:12,vandokkum:15}.
If these interpretations are correct, then of the galaxies discussed here, 
roughly half are in the process of undergoing a compaction event and half 
are growing inside out. 
Statistics cannot be drawn from one galaxy, however,
that \gn\ does not show evidence for compaction could reflect the theoretical 
argument that compaction is more common in high mass galaxies at high 
redshift when gas surface densities were higher. 

One important note here is that given the high dust columns toward the centers
of the massive $z\sim2$ galaxies implied by the submm data,
the expected dust attenuation is very high. 
Consequently, the central stellar mass surface density inferred based 
on rest-UV/optical data may be too low, meaning 
 the stellar mass effective radii may also be smaller than measured. 
If this were in fact the case, the effect in the top right panel of Fig.\,\ref{fig:ellipcomp}
would be to shift points to the left: a higher fraction of galaxies may in reality have
$r_e(submm)>r_e(M_*)$ and be growing inside-out. 

\section{Summary}
In this paper we investigate dust-obscured bulge growth in an \prog\ at $z=1.25$. 
We combine new millimeter dust continuum mapping from the NOEMA
interferometer with \ha, UV, and stellar mass maps to place constraints on
the formation pathways for bulge-disk systems. 

\gn\ displays a ring in \ha\ and UV emission, implying the central star formation is
strongly centrally suppressed in an absolute sense. 
However, when imaged at millimeter wavelengths, we instead
see centrally concentrated dust-continuum emission, meaning
this ring in unobscured star formation is likely filled in by dust-obscured star formation. 
This suggests that in this galaxy, the ring observed in \ha\ and UV
emission is caused by dust-obscuration rather than centrally suppressed
star formation. This is the main result of this paper: the bulge of this galaxy is building by 
dust-obscured star formation. 

To quantify this bulge building, we determine what fraction of the bulge growth underway
at $z\sim1$ can be accounted for by the star formation we observe. 
In \S\,\ref{section:csdev} we derive the quantity of bulge growth relative to disk growth
that would be required in order for a galaxy to remain on the scaling relation 
between central stellar mass surface density (as a proxy for bulge mass) 
and total stellar mass. We find that the dust-obscured star formation we
observe would move this galaxy along a trajectory with the same slope 
as this relation. Within the (significant) errors on this measurement, 
in this galaxy at this epoch, bulge growth can be explained by 
dust-obscured star formation. 
This galaxy lies above the main sequence and its optical morphology
suggests perhaps that it is undergoing a minor merger and/or have a large
clump in the disk. 
Taken together, our observations are consistent with a picture in which 
merging and disk instabilities drive gas to the center of the galaxy, 
boosting the global star formation rate, and building the bulge. 

The bulge and disk of this galaxy are building simultaneously. 
Although the bulge growth is dust-obscured,
the disk growth is apparent in the \restlam, \ha, and UV. 
Furthermore, while the dust-obscured star formation is more concentrated
than the un-obscured star formation, it has a similar or larger size than the 
stellar mass. The star formation we observe, although the errors are 
large, is consistent with the expected size evolution of Andromeda progenitors
at this epoch, gradually growing larger at a rate of $\Delta r_e/\Delta M_* \sim0.3 $.

This is in contrast to the dust continuum measurements of some massive galaxies at 
$z\sim2$, which are the putative progenitors of local massive elliptical galaxies. 
While mm/submm observations reveal dust-obscured bulge/dense core growth
for both Andromeda- and massive elliptical progenitors, some of the very massive 
galaxies at $z\sim2$ may actually be getting more compact due to star formation. 
It is expected that this strong form of compaction is more common at $z=2$ than
$z=1$ due to higher gas fractions and merger rates. However, given the extremely
high dust column densities toward the centers of these galaxies, we may be 
significantly underestimating the central stellar mass surface density already present.
If this were the case,  these galaxies would not in fact be undergoing
compaction in the sense of physically shrinking in radius.
Spatially resolved stellar population synthesis modeling which takes into account 
resolved millimeter constraints is needed before this can be definitively answered. 

The ability to map the structural growth of galaxies provides powerful constraints on 
the physical drivers of their evolution.
This requires maps of star formation and stellar mass which account for the
effects of dust and age. 
In this paper we showed that dust-obscured star formation can play
a key role in our understanding of the structural evolution of galaxies. 
Looking ahead, to really determine how galaxies are growing and 
whether dust-obscured star formation is responsible
for the building of bulges and dense cores requires dust continuum mapping 
with high spatial resolution statistical samples of galaxies at a range 
of redshifts beyond $z>1$ and across the SFR-\m\ plane. 
Because the progenitors of galaxies like the Milky Way and Andromeda
have lower masses, and correspondingly lower gas masses and metallicities,
longer integration times will be needed to map them -- but these 
measurements are key to understand how bulges build.
Additionally, high resolution mapping of molecular gas kinematics
will be essential to placing more stringent constraints on the physical processes 
responsible for building the dense centers of galaxies.



\end{document}